\documentclass[11ps,twoside]{article}

\usepackage{asp2004}
\usepackage{epsfig}
\usepackage{psfig}
\usepackage{natbib}
\begin{document}
\title{Radio emission from WR140}
\author{S.M. Dougherty}
\affil{National Research Council, D.R.A.O, P.O. Box 248, Penticton, BC, Canada}
\author{A.J. Beasley}
\affil{NRAO-AUI (ALMA), Camino del Observatorio 1515, \\ Las Condes, Santiago, Chile}
\author{M.J. Claussen}
\affil{NRAO, 1003 Lopezville Rd., Socorro NM 87801, USA}
\author{B.A. Zauderer}
\affil{Deptartment of Astronomy, University of Maryland, \\
College Park, MD 20742, USA}
\author{N.J. Bolingbroke}
\affil{Department of Physics and Astronomy, University of Victoria, \\
3800 Finnerty Rd, Victoria, BC, Canada}

\begin{abstract}
Milliarcsecond resolution Very Long Baseline Array (VLBA)
observations of the archetype WR+O star colliding-wind binary (CWB)
system WR\thinspace140 have been obtained at 23 epochs between orbital
phases 0.74 to 0.97. The emission in the wind-collision region (WCR)
is resolved as a bow-shaped arc which rotates as the orbital phase
progresses. This rotation provides for the first time the inclination of
the orbit ($122\deg\pm5\deg$), the longitude of the ascending node
($353\deg\pm3\deg$), and the orbit semi-major axis
($9.0\pm0.5$~mas). The implied distance is $1.85\pm0.16$~kpc, which
requires the O star to be a supergiant, and leads to a wind-momentum
ratio of 0.22. Quasi-simultaneous Very Large Array (VLA) observations
show the synchrotron spectra evolve dramatically through the orbital
phases observed, exhibiting both optically thin and optically thick
emission.  The optically thin emission maintains a spectral index of
$-0.5$, as expected from diffusive shock acceleration.
\end{abstract}

The archetype of CWB systems is the 7.9-year period WR+O system
WR\thinspace140 (HD\thinspace193793).  Its highly eccentric orbit
($e\approx0.88$) modulates the dramatic variations in the emission
from the system observed at many wavelengths.  At radio wavelengths
there is a slow rise from a low thermal state close to periastron of a
few mJy, to a frequency-dependent peak in emission of 10's of mJy
between orbital phase 0.65 to 0.85, before a precipitous decline just
before periastron. The radio variations have been widely attributed
to an underlying synchrotron source viewed through the changing
free-free opacity of the extended stellar winds of the
binary system along the line-of-sight to the WCR as the orbit
progresses \citep{Williams:1990, White:1995}. However, none of the
free-free opacity models explain the radio light curve in a
satisfactory manner.  Models that include processes intrinsic to the
WCR are now being explored \citetext{Pittard et al., in preparation}.

We report briefly on high resolution observations of WR\thinspace140
obtained with the VLBA that image structures in the WCR at a linear
resolution of a few AU, approximately the stellar separation at
periastron, along with quasi-simultaneous VLA observations.
\citet{Dougherty:2004} include a more detailed description of this
work.

\section{Observations}
\begin{figure*}
\vspace{3.8cm} 
\includegraphics{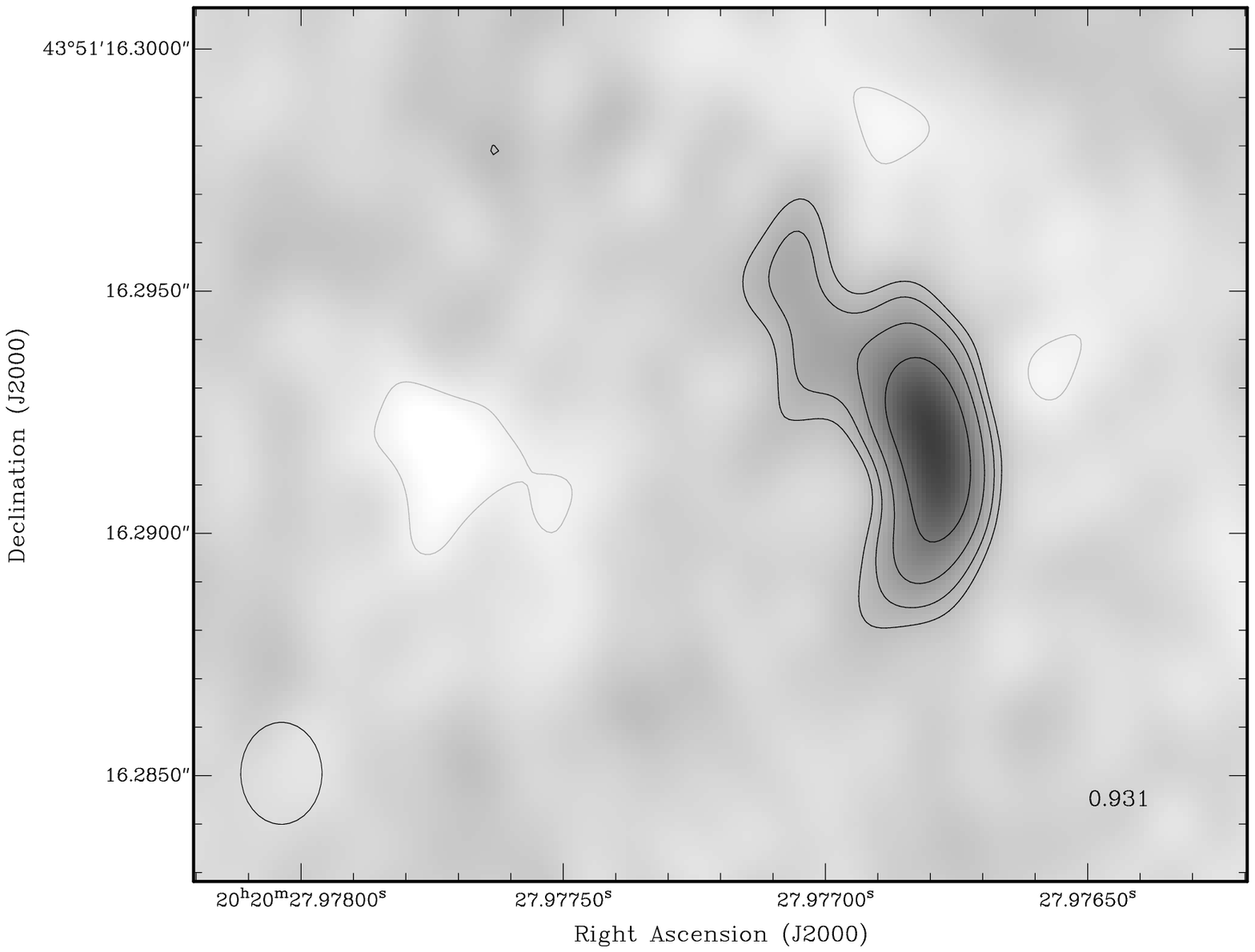}
\includegraphics{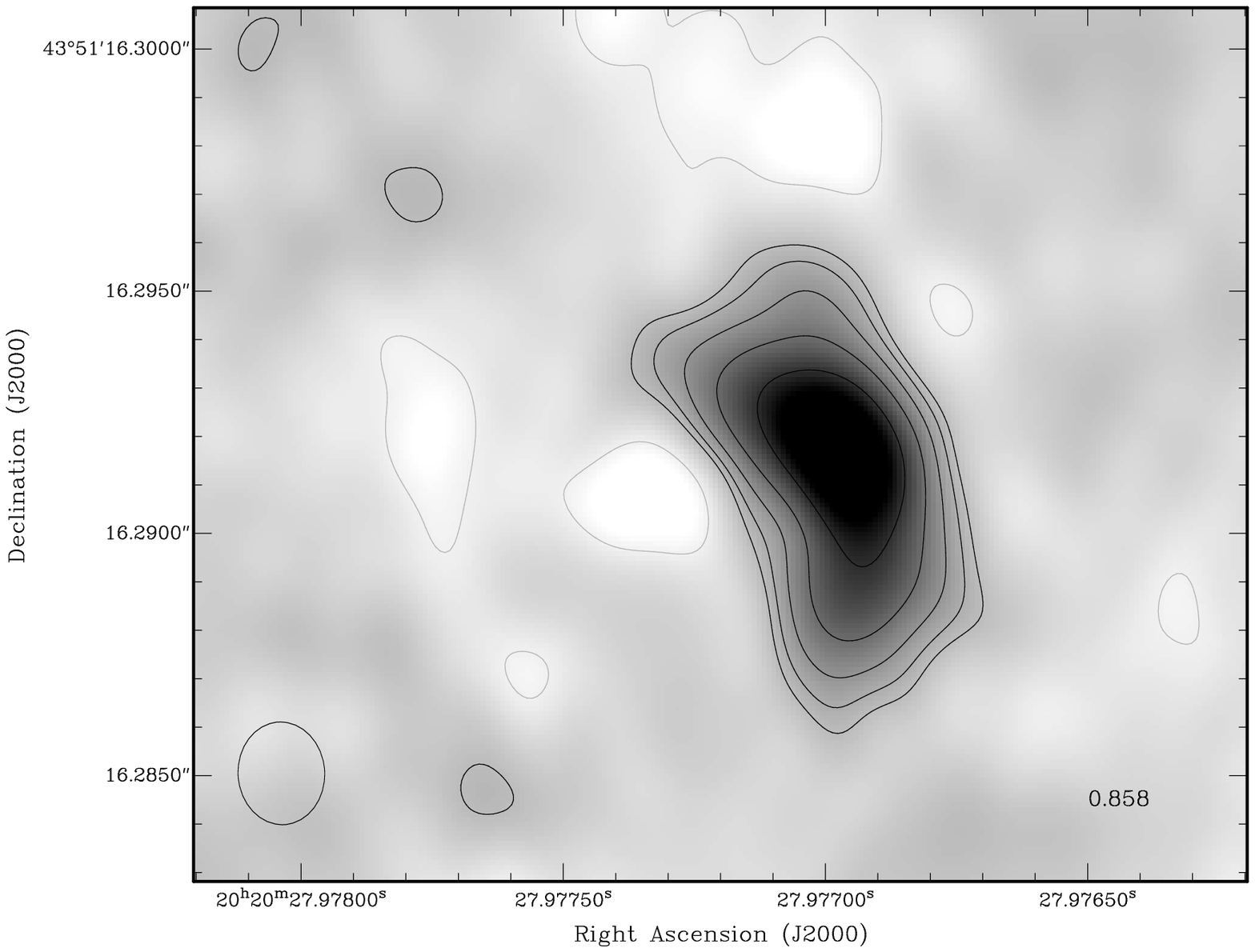}
\includegraphics{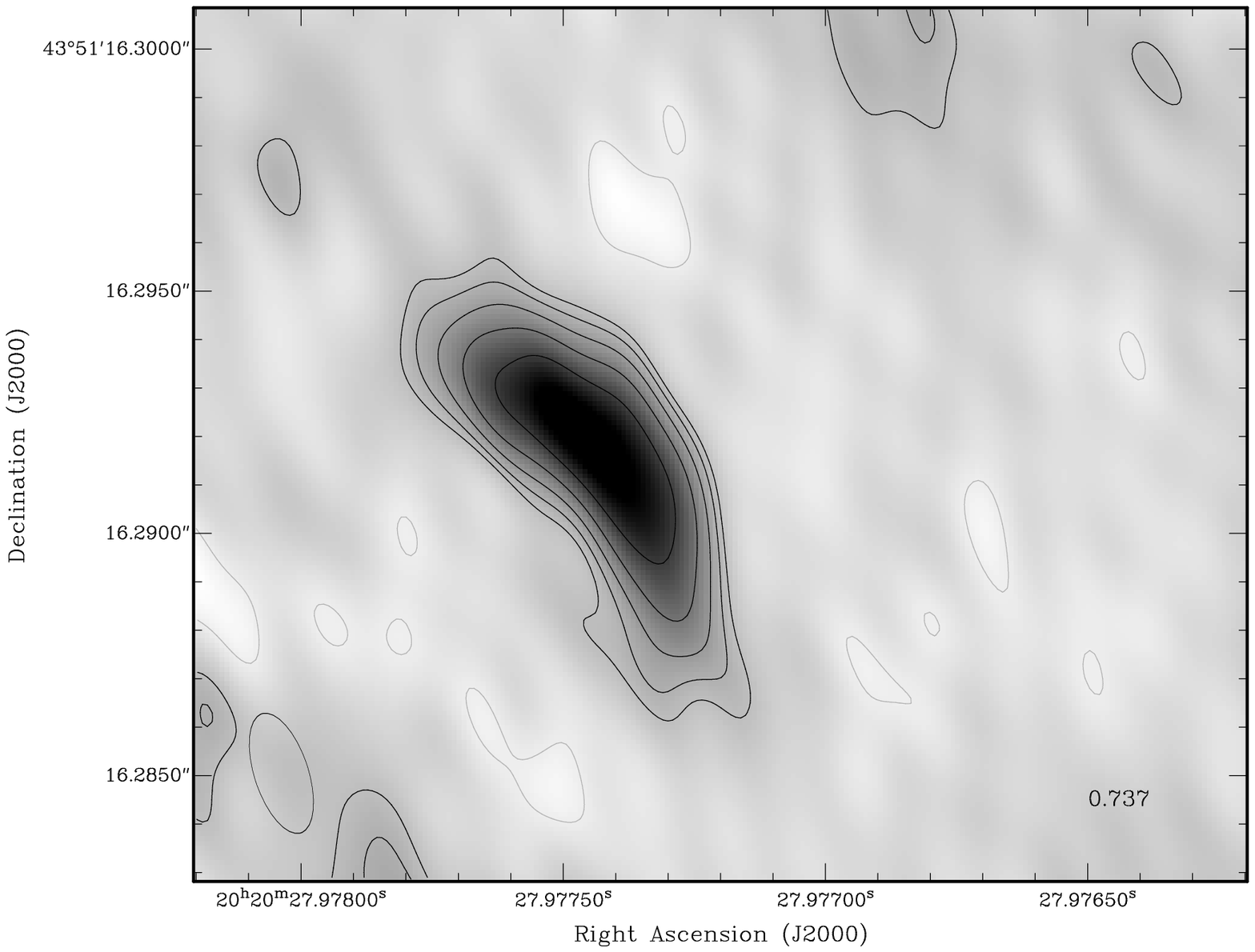}
\caption{\footnotesize 8.4-GHz VLBA observations of WR140 at orbital
phases 0.74, 0.86 and 0.93. The contour levels are -1, 1, 1.6, 2.6,
4.1, 6.6, 10.5$\rho$ where $\rho=220~\mu$Jy~beam$^{-1}$. The contour levels
and greyscale are identical in each image. The beam size is shown in
the lower left corner of the images, and is typically
$2.0\times1.5$~mas$^2$, which at a distance of 1.5 kpc gives a linear
resolution of 3 AU. Rotation and proper motion of the WCR are
clear. \label{fig:xband_vlba}}
\end{figure*}

Observations of WR\thinspace140 were obtained at 23 epochs using the
VLBA. The campaign started on Jan 4, 1999 near the peak of radio
emission around orbital phase 0.75 and was completed Nov 18, 2000 when
the radio emission had declined to its low level.  The resulting
phase-referenced 8.4-GHz images at three of the observed epochs are
shown in Fig.~\ref{fig:xband_vlba}. Additionally, we also obtained
closely concurrent observations with the VLA at five frequencies
between 1.4 and 22 GHz.

The 8.4-GHz emission detected by the VLBA is clearly resolved.  We
identify this emission as arising from the WCR in WR\thinspace140
since this is the only emission in the system with sufficient
brightness temperature ($>10^{5}$~K) to be detected by the VLBA.  The
stellar winds, with brightness of $\sim10^4$~K are undetected.  A
bow-shaped ridge of emission is observed at most epochs, as
anticipated for the WCR from model calculations
\citep[e.g. see][]{Eichler:1993, Canto:1996, Dougherty:2003}, with the
bow shock wrapping around the star with the lower wind momentum -
typically the O star. Between orbit phase 0.74 and 0.95, the WCR
exhibits rotation from ``pointing'' NW to W, in addition to an east to
west proper motion of $\sim10$~mas.

\section{Orbital parameters of WR\thinspace140}
\label{sec:orbit}
Many of the orbital parameters in WR\thinspace140 are well-determined
from radial velocity measurements \citep[see][and references
therein]{Marchenko:2003}. However, the orbital inclination ($i$),
semi-major axis ($a$) and the longitude of the ascending node
($\Omega$) can only be determined if the system can be resolved into a
``visual'' binary around the orbit. The two stellar components in
WR\thinspace140 have been resolved using the Infrared-Optical
Telescope Array (IOTA) interferometer on June 17, 2003 to have a
separation of $12.9^{+0.5}_{-0.4}$~mas at a position angle of
${151.7^{+1.8}_{-1.3}}$~degrees east of north \citep{Monnier:2004}.
Assuming $P=2899$~days, $T_o=2446147.4$, $e=0.881$ and $\omega=47\deg$
\citep{Marchenko:2003}, this single epoch observation at orbital phase
0.297 gives families of possible solutions for ($i,\Omega,a$).

\begin{figure}
\vspace{3.8cm}
\includegraphics{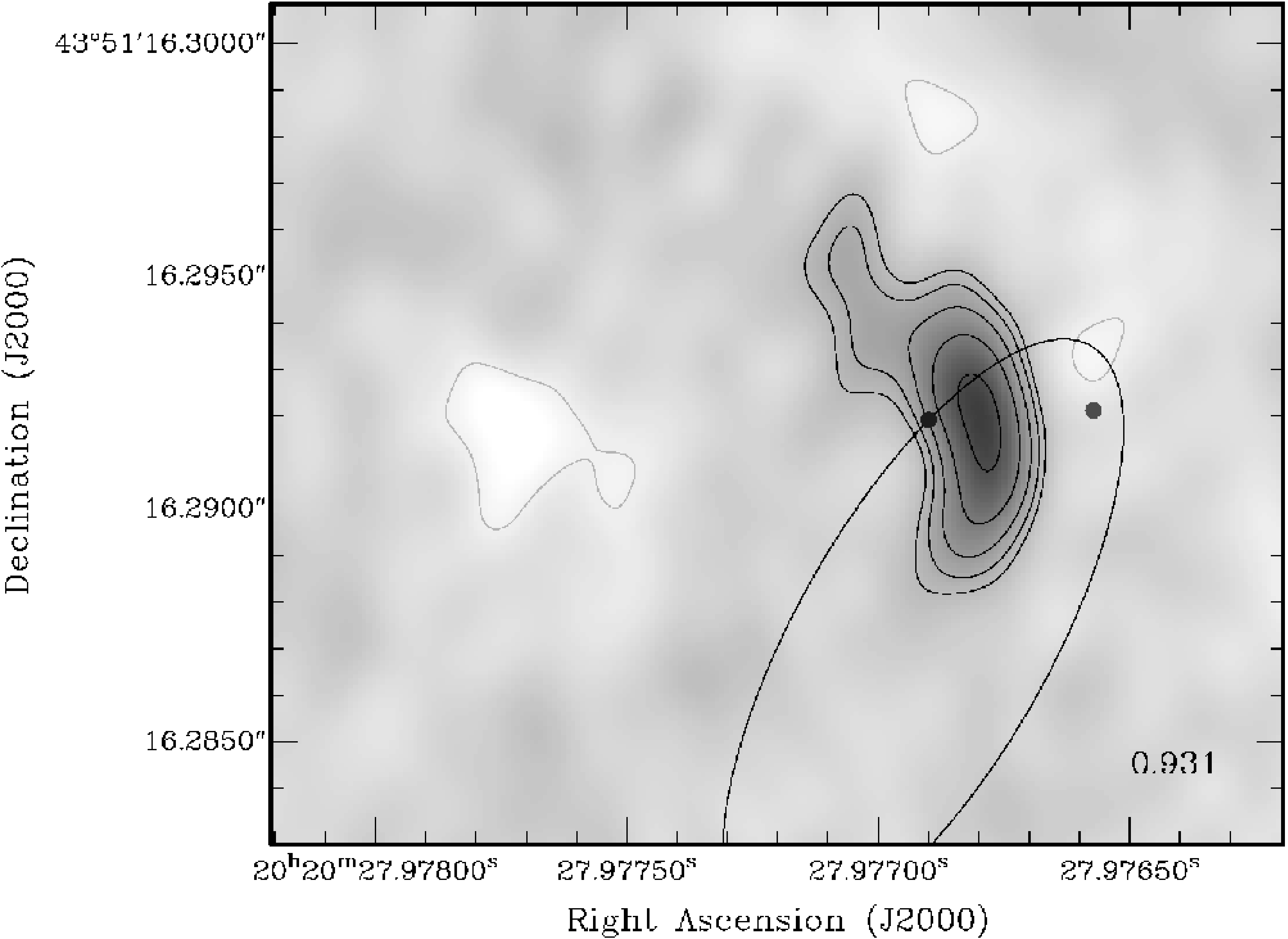}
\includegraphics{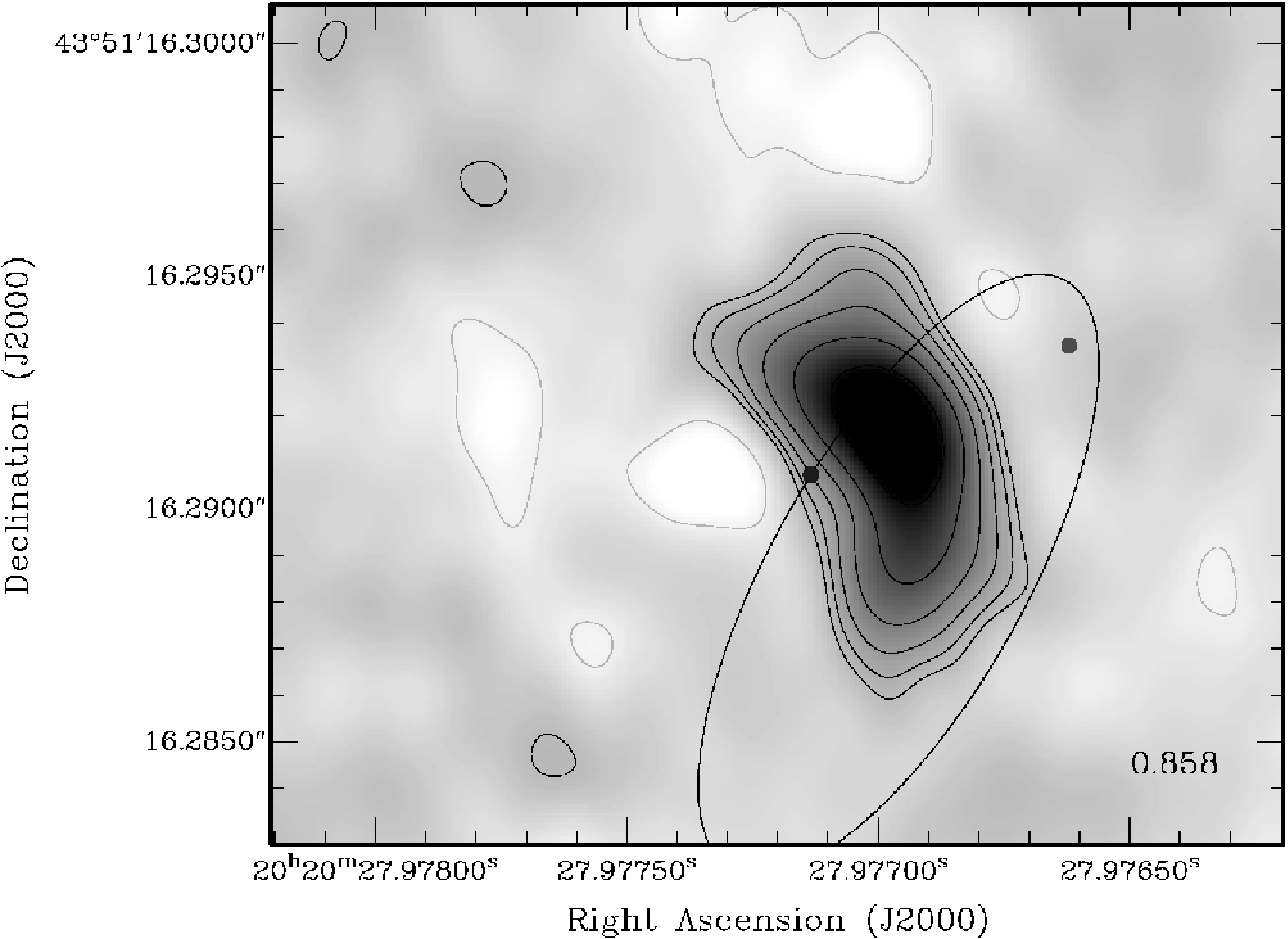}
\includegraphics{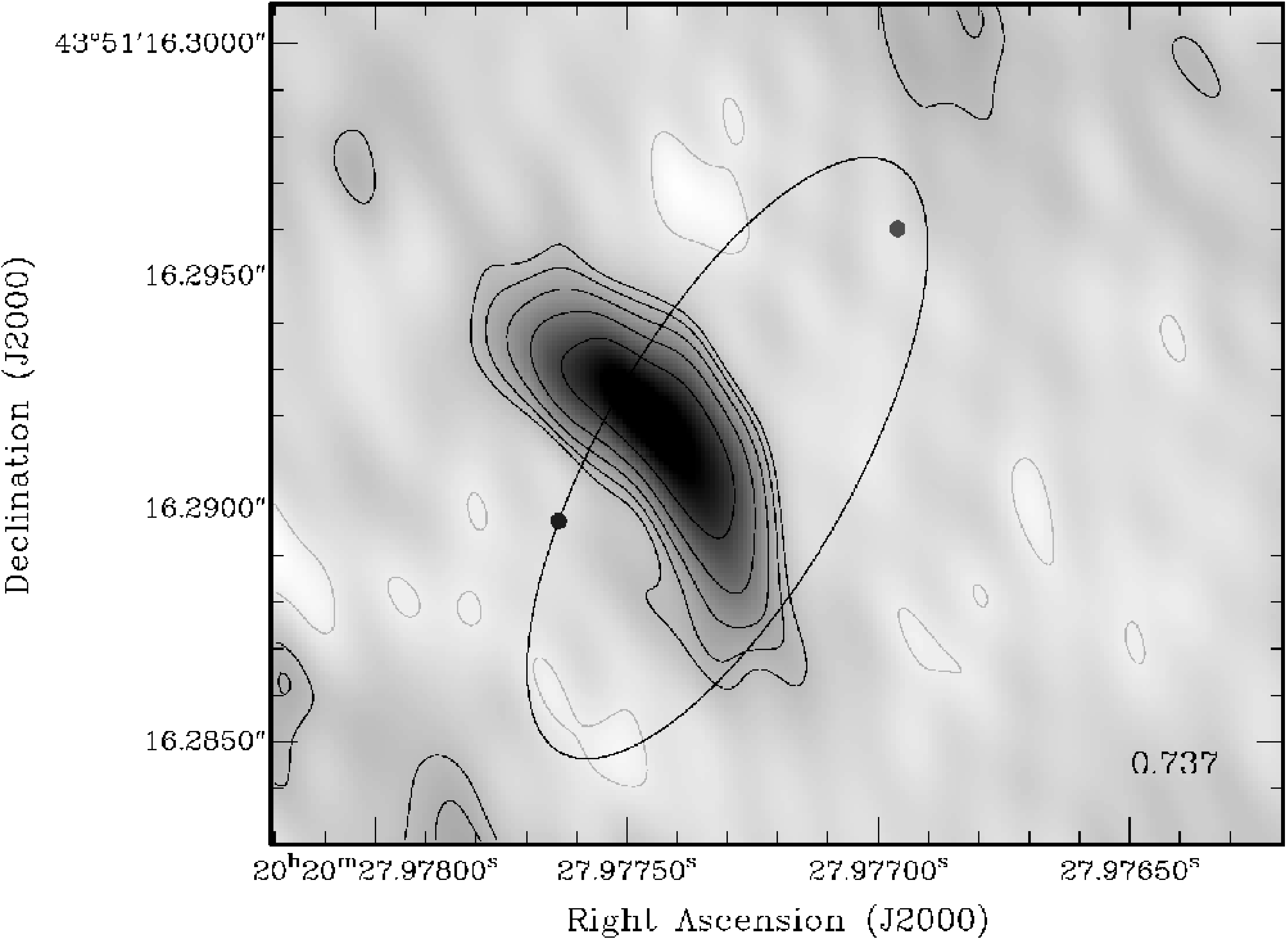}
\caption[]{The derived orbit of WR\thinspace140 on the plane of the
sky using $e=0.88$, $\omega=47\deg$, $\Omega=353\deg$, $i=122\deg$ and
$a=9.0$~mas at orbital phase 0.737, 0.858 and 0.931, overlaid on the
VLBA 8.4 GHz images. The WR star is to the W (right) of the WCR at
these phases.  The rotation of the WCR as the orbit progresses is
clear. The relative position of the stars to the WCR was determined
using a wind momentum ratio of $0.22$ (see
\textsection{~\ref{sec:distance}})
\label{fig:orbit_vlba}}
\end{figure}

Currently, the VLBA observations of the WCR are the only means to
determine uniquely $i$, and hence $\Omega$ and $a$, from the
possible IOTA solutions. Under the assumption that the free-free
opacity along the line-of-sight to the WCR is sufficiently low as to
not impact the apparent distribution of emission from the WCR, we
expect the ``arc'' of WCR emission to wrap around the star with the
lower wind momentum - the O star.  In this case, the rotation of the
WCR as the orbit progresses implies that O star moves from the SE to
close to due E of the WR star over the period of the VLBA
observations.  Also, if it is assumed the axis of symmetry of the
emission from the WCR is coincident with the projection on the plane
of the sky of the line-of-centres of the two stars, we can derive the
orbital inclination from the change in the orientation of the WCR with
orbital phase.  Each ($i,\Omega$) family provides a unique set of
position angles for the projected line-of-centres as a function of
orbital phase.  By fitting the position angle of the line of symmetry
of the WCR as a function of orbit phase for different sets of
($i,\Omega$), we find a best-fit solution of $i=122\deg\pm5\deg$ and
$\Omega=353\deg\pm3\deg$. These values lead to a semi-major axis of
$a=9.0\pm0.5$~mas, and a projected orbit that evolves as shown in
Fig.~\ref{fig:orbit_vlba}.

The derived orbit inclination is consistent with values previously
suggested. However, it presents a challenge to current models of dust
formation. To date, most models of dust formation in WR\thinspace140
assume that the gas from which dust is formed in the WCR is compressed
within $\sim0.15$~yr of periastron passage \citep{Williams:1990}. The
subsequent motion of the compressed gas is determined by the velocity
of this material when it is compressed.  Since the momentum of the WR
star wind is higher than that of the O star, this material moves away
from the WR star, along the WCR.  With the orbit orientation derived
here, the O star is NW of the WR star during periastron, and material
compressed at periastron will therefore have a proper motion to the
NW. However, high-resolution IR observations show that dust ejected
during the 2001 periastron passage has proper motion to the south and
east, away from the WR star \citep{Monnier:2002}. New dust models are
now attempting to address this challenge \citetext{Williams, this
volume}.

\section{Basic system parameters of WR\thinspace140}
\label{sec:distance}
Distance estimates of WR stars are typically based on absolute
magnitude calibrations that often have large scatter. Having
determined the orbital inclination and semi-major axis it is now
possible to make an estimate of the distance to WR\thinspace140 {\em
independent of any stellar parameters}. \citet{Marchenko:2003}
determined $a\,\sin i = 14.10\pm0.54$~AU, which leads to $a=16.6\pm1.1$~AU for
$i=122\deg\pm5\deg$. Along with the derived semi-major axis of
$a=9.0\pm0.5$~mas, these give a distance of $1.85\pm0.16$~kpc.

\begin{table}[t]
\caption{Basic parameters of WR\thinspace140\label{tab:stellar_parms}}
\begin{tabular}{llll}
\hline
Parameter & Primary & Secondary & System \\
\hline
Distance (kpc) & & & 1.85 \\
M$_v$ & -6.4 & -5.6 & -6.8\\
Spectral type & 0.4-5 I $^{a}$ & WC7 & \\
BC    & -4.3 $^a$ & -3.4 $^{b}$ & \\
M$_{\rm bol}$ & -10.7 & -9.0 & \\
log (L$_{\rm bol}$/L$_\odot$) & 6.18 $^{c}$ & 5.5 $^{c}$ & \\
Mass (M$_\odot$) & $54\pm10$ & $20\pm4$ \\
v$_\infty$ (km s$^{-1}$) & 3100 $^{d}$ & 2860 $^{e}$\\
$\dot{\rm M}$ (${\rm M}_\odot$~yr$^{-1}$) & $8.7\times10^{-6}$ & $4.3\times10^{-5}$& \\
\hline
\end{tabular}

\scriptsize
$^{a}${Based on M$_v$ and the calibration of \citet{Vacca:1996}}
$^{b}${\citet{Williams:1990}}
$^{c}${Calculated assuming M$_{\odot, bol}=4.75^m$ \citep{Allen:1976}}
$^{d}${\citet{SetiaGunawan:2001}}
$^{e}${\citet{Eenens:1994}}
\end{table}

This distance is somewhat larger than the usually quoted value of
1.3~kpc deduced by \citet{Williams:1990} from the luminosity of the
system. Since the primary O-star luminosity indicator is masked by the
WC7 spectrum, \citet{Williams:1990} assumed a main sequence O4-5 star
with an absolute magnitude of $-5.6$ and took that of the WC7 star to
be $-4.8$. With the system at $1.85$~kpc, the absolute magnitude of
the O4-5 star becomes $-6.4$, suggesting it is a supergiant
\citep[see][Table 7] {Vacca:1996}.

With the increase in distance, a reassessment of the mass-loss rates
of the two stars is appropriate.  Based on the X-ray luminosity
measured by ASCA \citep{Zhekov:2000}, the mass-loss rate for the WC
star at 1.85 kpc is $4.3\times10^{-5}~{\rm M}_\odot$~yr$^{-1}$.
\citet{Repolust:2004} suggests values of $8.6-8.8\times10^{-6}~{\rm
M}_\odot$~yr$^{-1}$ for O4-5 supergiants.  Along with the wind speeds
(Table~\ref{tab:stellar_parms}) these mass-loss rates imply a wind
momentum ratio $\eta=0.22$.  This wind-momentum ratio is considerably
higher than the 0.035 deduced by \citet{Williams:1990}. The higher
value of $\eta$ derived here, however, implies a half-opening angle of
the WCR of $63\deg$ (following \citet{Eichler:1993}), consistent with
$65\deg\pm10\deg$ derived from these VLBA observations.

\section{The radio spectra of WR\thinspace140}
\begin{figure}[t]
\begin{center}
\includegraphics[angle=0,scale=0.6]{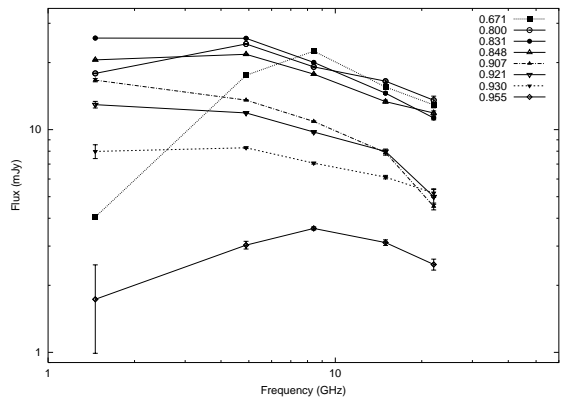}
\caption[]{Synchrotron spectra of WR\thinspace140 at several orbital
phases, determined by subtracting the thermal spectrum at phase 0.974
from the spectra at each phase.  The key to the plot is shown in the
upper right-hand corner.\label{fig:sync_spectra}}
\end{center}
\end{figure}

The new VLA data at five frequencies allow us to observe the radio
spectrum and its evolution better than previously possible, most
particularly at the higher frequencies.  At phase 0.974, the spectrum
is a power-law with a spectral index of $0.72\pm0.03$, a value
characteristic of the stellar winds in WR+OB binary
systems. Furthermore, the flux levels at this phase are consistent
with the wind densities implied by the parameters in
Table~\ref{tab:stellar_parms}.  Assuming the thermal emission from
WR\thinspace140 is essentially constant throughout the orbit, the
synchrotron spectra at each observed phase can be determined by simply
subtracting the thermal flux at phase 0.974 from the total flux. The
resulting spectra are shown in Fig.~\ref{fig:sync_spectra}.

The synchrotron spectra between phases 0.67 and 0.92 are optically
thin at several frequencies, with a spectral index that appears to be
closely constant, with a slope of $-0.5\pm0.1$, as expected for
diffusive shock acceleration of electrons in strong, non-relativistic
shocks \citep[e.g.][and references therein]{Bell:1978, Drury:1983, Jones:1991}. The optically
thick spectrum apparent during the bulk of the orbit has been widely
attributed to free-free absorption in the stellar winds along the
line-of-sight to the WCR. Unfortunately, these models are too simple
to explain the radio observations of WR\thinspace140, as readily
acknowledged by their authors. The VLBA observations
(Fig.~\ref{fig:xband_vlba}) show the WCR as a distributed emission
region and the lines-of-sight to the WCR traverse different regions of
the stellar winds. As a result, the emerging emission will be a
combination of both optically thick and thin emission since even
though lines-of-sight to the apex may be optically thick, a
substantial amount of emission arises from optically thin
lines-of-sight to the downstream flow. Using the newly derived orbit
and assuming a half-opening angle for the WCR of $63\deg$, we now know
the lines-of-sight to the WCR traverse the O-star wind between orbit
phases 0.24 and 0.99 during which the most dramatic changes in the
radio emission are observed. Clearly, if stellar wind free-free
opacity plays any role in determining the observed spectra, it is the
O-star wind opacity that is of most concern, not that of the WR star.
Another shortcoming of previous models is the assumption that
synchrotron emission is optically thin at all observing frequencies,
and at all orbital phases. The optically thick component of the
spectra may, at least in part, be due to mechanisms intrinsic to the
WCR.

New radiative transfer models, based on a fully consistent
hydrodynamic treatment of the WCR, have started to explore the impact
of a number of processes on the radio emission from CWBs, including
free-free opacity in the stellar winds and the WCR, synchrotron
self-absorption, Coulombic cooling through interactions with
post-shock ions, plasma effects such as the Razin effect, and Inverse
Compton cooling by the intense ultra-violet radiation field of the
nearby massive stars. These models have been very successful in
explaining the radio emission from very wide CWBs such as WR\thinspace
147 \citep{Dougherty:2003}, and are now maturing to the point where
they will provide more insight to the mechanisms acting in
WR\thinspace140 \citetext{Pittard et al., in preparation}. The
observations presented here represent the essential constraints for
these new models.

\end{document}